\documentclass[preprint,12pt]{elsarticle}
\usepackage[margin=1in]{geometry}

\usepackage{mathtools}
\usepackage{amssymb}
\usepackage{amsfonts}
\usepackage{lipsum}
\usepackage{graphicx}
\usepackage{epstopdf}
\usepackage{hyperref}
\usepackage[export]{adjustbox}
\usepackage{float}

\usepackage{xcolor}
\definecolor{mintedbg}{gray}{0.95}

\usepackage{amsmath}

\DeclareMathOperator*{\argmin}{arg\,min}

\usepackage{mdframed}

\DeclareFixedFont{\ttb}{T1}{txtt}{bx}{n}{12} 
\DeclareFixedFont{\ttm}{T1}{txtt}{m}{n}{12}  

\usepackage{color}
\definecolor{deepblue}{rgb}{0,0,0.5}
\definecolor{deepred}{rgb}{0.6,0,0}
\definecolor{deepgreen}{rgb}{0,0.5,0}

\usepackage{listings}

\newcommand\pythonstyle{
    \lstset{
        language=Python,
        basicstyle=\small\ttfamily,
        numberstyle=\color{gray},
        stringstyle=\color[HTML]{933797},
        commentstyle=\color[HTML]{228B22}\small\sffamily,
        emph={[2]from,import,pass,return}, emphstyle={[2]\color[HTML]{DD52F0}},
        emph={[3]range}, emphstyle={[3]\color[HTML]{D17032}},
        emph={[4]for,in,def}, emphstyle={[4]\color{blue}},
        showstringspaces=false,
        breaklines=true,
        prebreak=\mbox{{\color{gray}\tiny$\searrow$}},
        numbers=left,
        xleftmargin=0pt
    }
}

\lstnewenvironment{python}[1][]
{
\pythonstyle
\lstset{#1}
}
{}

\newcommand\pythoninline[1]{{\pythonstyle\lstinline!#1!}}

\newcommand\bashstyle{\lstset{
language=Bash,
basicstyle=\ttm\small,
otherkeywords={self},             
keywordstyle=\ttm,
showstringspaces=false            %
}}

\lstnewenvironment{bash}[1][]
{
\bashstyle
\lstset{#1}
}
{}

\begin{document}

\begin{frontmatter}

\title{SciANN: A Keras/Tensorflow wrapper for scientific computations and physics-informed deep learning using artificial neural networks}

\author[MIT]{Ehsan Haghighat}
\author[MIT]{Ruben Juanes}
\address[MIT]{Massachusetts Institute of Technology, Cambridge, MA}

\begin{abstract}
  In this paper, we introduce SciANN, a Python package for scientific computing and physics-informed deep learning using artificial neural networks. SciANN uses the widely used deep-learning packages Tensorflow and Keras to build deep neural networks and optimization models, thus inheriting many of Keras's functionalities, such as batch optimization and model reuse for transfer learning. SciANN is designed to abstract neural network construction for scientific computations and solution and discovery of partial differential equations (PDE) using the physics-informed neural networks (PINN) architecture, therefore providing the flexibility to set up complex functional forms. We illustrate, in a series of examples, how the framework can be used for curve fitting on discrete data, and for solution and discovery of PDEs in strong and weak forms. We summarize the features currently available in SciANN, and also outline ongoing and future developments. 
\end{abstract}




\end{frontmatter}


\section{Introduction}
Over the past decade, artificial neural networks, also known as deep learning, have revolutionized many computational tasks, including image classification and computer vision~\cite{Bishop2006,NIPS2012_4824,Lecun2015}, search engines and recommender systems~\cite{jannach2010recommender, zhang2019deep}, speech recognition~\cite{graves2013speech}, autonomous driving~\cite{bojarski2016end}, and healthcare~\cite{miotto2018deep} (for a review, see, e.g.~\cite{Goodfellow2016}). Even more recently, this data-driven framework has made inroads in engineering and scientific applications, such as earthquake detection~\cite{Kong2018,Ross2019,Bergen2019}, fluid mechanics and turbulence modeling~\cite{Brenner2019,Brunton2020}, dynamical systems~\cite{Dana2020}, and constitutive modeling~\cite{Tartakovsky2018, Xu2020}. A recent class of deep learning known as physics-informed neural networks (PINN) \cite{Raissi2019}, where the network is trained simultaneously on both data and the governing differential equations, has been shown to be particularly well suited for solution and inversion of equations governing physical systems, in domains such as fluid mechanics~\cite{Raissi2019, Raissi2018d}, solid mechanics~\cite{Haghighat2020} and dynamical systems~\cite{Rudy2019}. This increased interest in engineering and science is due to the increased availability of data and open-source platforms such as Theano~\cite{bergstra2010theano}, Tensorflow~\cite{abadi2016tensorflow}, MXNET~\cite{chen2015mxnet}, and Keras~\cite{chollet2015keras}, which offer features such as high-performance computing and automatic differentiation~\cite{Gune2018}. 

Advances in deep learning have led to the emergence of different neural network architectures, including densely connected multi-layer deep neural networks (DNNs), convolutional neural networks (CNNs), recurrent neural networks (RNNs) and residual networks (ResNets). This proliferation of network architectures, and the (often steep) learning curve for each package, makes it challenging for new researchers in the field to use deep learning tools in their computational workflows. In this paper, we introduce an open-source Python package, SciANN, developed on Tensorflow and Keras, which is designed with scientific computations and physics-informed deep learning in mind. As such, the abstractions used in this programming interface target engineering applications such as model fitting, solution of ordinary and partial differential equations, and model inversion (parameter identification).

The outline of the paper is as follows. We first describe the functional form associated with deep neural networks. We then discuss different interfaces in SciANN that can be used to set up neural networks and optimization problems. We then illustrate SciANN's application to curve fitting, the solution of the Burgers equation, and the identification of the Navier--Stokes equations and the von~Mises plasticity model from data. Lastly, we show how to use SciANN in the context of the variational PINN framework~\cite{Kharazmi2020}. The examples discussed here and several additional applications are freely available at \href{https://github.com/sciann/sciann-applications}{github.com/sciann/sciann-applications}.


\section{Artificial Neural Networks as Universal Approximators}
\label{sec:DNN}

A single-layer feed-forward neural network with inputs $\mathbf{x} \in \mathbb{R}^m$, outputs $\mathbf{y} \in \mathbb{R}^n$, and $d$~hidden units is constructed as:
\begin{equation}\label{eq:dnn1}
    \mathbf{y} = \mathbf{W}^1 \sigma(\mathbf{W}^0 \mathbf{x} + \mathbf{b}^0) + \mathbf{b}^1,
\end{equation}
where ($\mathbf{W}^0 \in \mathbb{R}^{d \times m}$, $\mathbf{b}^0 \in \mathbb{R}^{d}$), ($\mathbf{W}^1 \in \mathbb{R}^{n \times d}$, $\mathbf{b}^1 \in \mathbb{R}^{n}$) are parameters of this transformation, also known as weights and biases, and $\sigma$~is the activation function. As shown in~\cite{Hornik1989, Cybenko1989}, this transformation can approximate any measurable function, independently of the size of input features~$m$ or the activation function~$\sigma$. If we define the transformation $\Sigma$ as $\Sigma^{i}(\hat{\mathbf{x}}^{i}) := \hat{\mathbf{y}}^{i} = \sigma^{i}(\mathbf{W}^{i} \hat{\mathbf{x}}^{i} + \mathbf{b}^{i})$ with $\hat{\mathbf{x}}^i$ as the input to and $\hat{\mathbf{y}}^i$ as the output of any hidden layer~$i$, $\mathbf{x}=\hat{\mathbf{x}}^0$ as the main input to the network, and $\mathbf{y} = \Sigma^L(\hat{\mathbf{x}}^{L})$ as the final output of the network, we can construct a general $L$-layer neural network as composition of $\Sigma^i$~functions as:
\begin{equation}\label{eq:dnn2}
    \mathbf{y} = \Sigma^L \circ \Sigma^{L-1} \circ \dots \circ \Sigma^0(\mathbf{x}),
\end{equation}
with $\sigma^i$ as activation functions that make the transformations nonlinear. Some common activation functions are:
\begin{equation} \label{eqn:dnn3}
\begin{split}
   \textrm{ReLU} &: \hat{x} \mapsto \hat{x}^+, \\
   \textrm{sigmoid} &: \hat{x} \mapsto 1/(1+e^{\hat{x}}), \\
   \tanh &: \hat{x} \mapsto (e^{\hat{x}} - e^{-\hat{x}}) / (e^{\hat{x}} + e^{-\hat{x}}).
\end{split}
\end{equation}
In general, this multilayer feed-forward neural network is capable of approximating functions to any desired accuracy~\cite{Hornik1989, Hornik1991}. Inaccurate approximation may arise due to lack of a deterministic relation between input and outputs, insufficient number of hidden units, inadequate training, or poor choice of the optimization algorithm.

The parameters of the neural network, $\mathbf{W}^i$ and $\mathbf{b}^i$ of all layers $i=\left\{ 0 \dots L \right\}$, are identified through minimization using a back-propagation algorithm~\cite{Rumelhart1986}. For instance, if we approximate a field variable such as temperature~$T$ with a multi-layer neural network as $T(\mathbf{x}) \approx \hat{T}(\mathbf{x}) = \mathcal{N}_T(\mathbf{x}; \mathbf{W}, \mathbf{b})$, we can set up the optimization problem as 
\begin{equation}\label{eq:dnn4}
    \argmin_{\mathbf{W}, \mathbf{b}} \mathcal{L}(\mathbf{W}, \mathbf{b}) := \left \| T(\mathbf{x^*}) - \hat{T}(\mathbf{x^*}) \right \| = \left \| T(\mathbf{x^*}) - \mathcal{N}_T(\mathbf{x^*}; \mathbf{W}, \mathbf{b}) \right \|,
\end{equation}
where $\mathbf{x}^*$~is the set of discrete training points, and $\left\|\circ\right\|_p$~is the mean squared norm. Note that one can use other choices for the loss function~$\mathcal{L}$, such as mean absolute error or cross-entropy. The optimization problem~\eqref{eq:dnn4} is nonconvex, which may require significant trial and error efforts to find an effective optimization algorithm and optimization parameters.

\begin{figure}[!ht]
  \centering
  \includegraphics[width=0.7\textwidth]{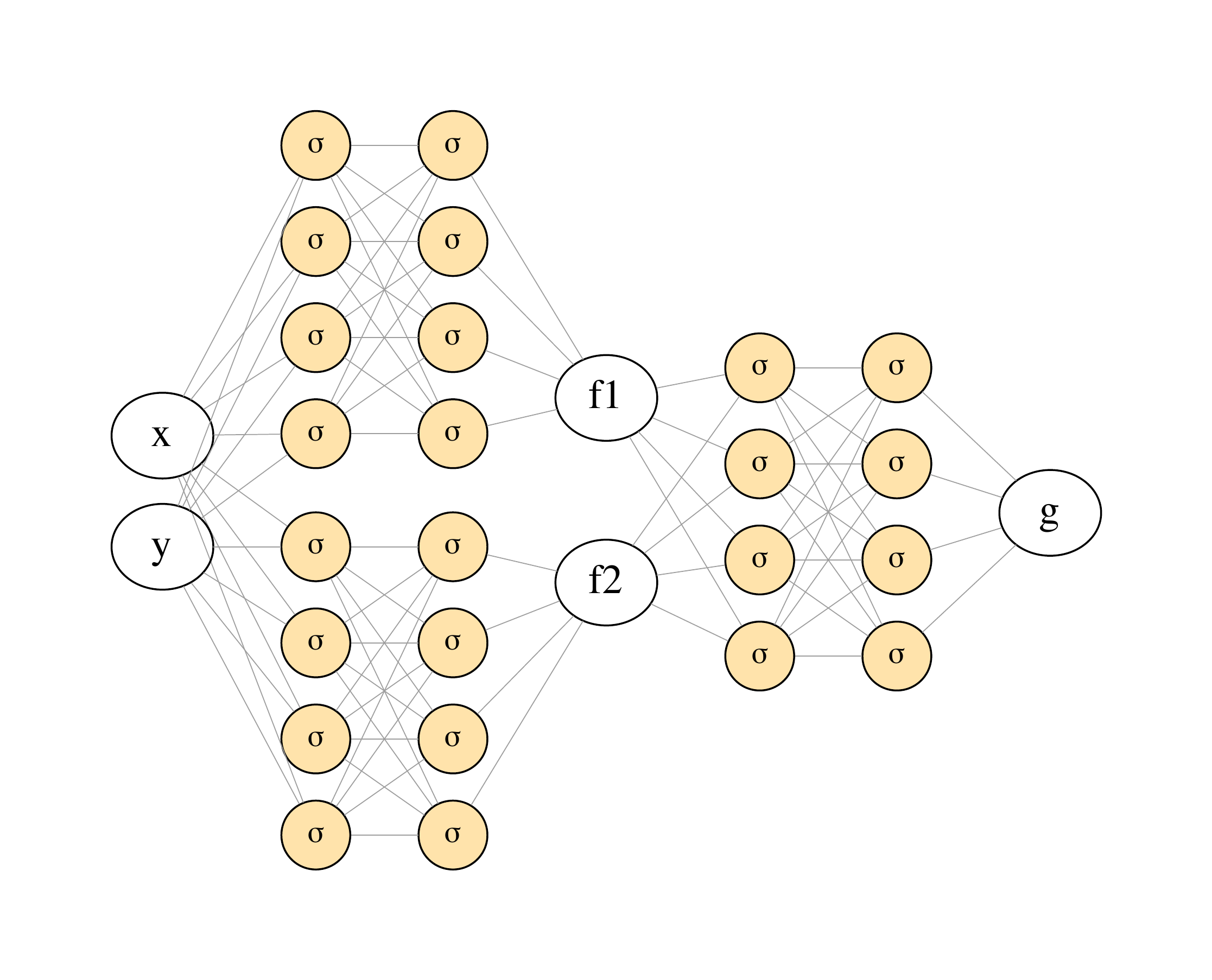}
  \caption{A sample multi-net architecture to construct a complex functional space~$g$ as $g(x,y) = g(f_1(x,y), f_2(x,y))$.}
  \label{fig:nn_multinet}
\end{figure}

We can construct deep neural networks with an arbitrary number of layers and neurons. We can also define multiple networks and combine them to generate the final output. There are many types of neural networks that have been optimized for specific tasks. An example is the ResNet architecture introduced for image classification, consisting of many blocks, each of the form: 

\begin{equation}\label{eq:dnn5}
	\mathbf{z}^k = \Sigma^{k2} \circ \Sigma^{k1} \circ \Sigma^{k0}(\mathbf{z}^{k-1}) + \mathbf{z}^{k-1},
\end{equation}
where $k$~is the block number and $\mathbf{z}^{k-1}$~is the output of previous block, with~$\mathbf{x}=\mathbf{z}^{0}$ and $\mathbf{y}=\mathbf{z}^{K}$ as the main inputs to and outputs of the network. Therefore, artificial neural networks offer a simple way of constructing very complex but dependent solution spaces (see, e.g., Fig.~\ref{fig:nn_multinet}). 

\section{SciANN: Scientific Computing with Artificial Neural Networks}

SciANN is an open-source neural-network library, based on Tensorflow~\cite{abadi2016tensorflow} and Keras~\cite{chollet2015keras}, which abstracts the application of deep learning for scientific computing purposes. In this section, we discuss abstraction choices for SciANN and illustrate how one can use it for scientific computations. 

\subsection{Brief description of SciANN}

SciANN is implemented on the most popular deep-learning packages, Tensorflow and Keras, and therefore it inherits all the functionalities they provide. Among those, the most important ones include graph-based automatic differentiation and massive heterogeneous high-performance computing capabilities. It is designed for an audience with a background in scientific computation or computational science and engineering. 

SciANN currently supports fully connected feed-forward deep neural networks, and recurrent networks are under development. Some architectures, such as convolutional networks, are not a good fit for scientific computing applications and therefore are not currently in our development plans. Tensorflow and Keras provide a wide range of features, including optimization algorithms, automatic differentiation, and model parameter exports for transfer learning. 

To install SciANN, one can simply use the Python's pip package installer as:
\begin{mdframed}[backgroundcolor=mintedbg, linecolor=mintedbg, innerleftmargin=0, innertopmargin=0,innerbottommargin=0]
\begin{bash}
    pip install sciann
\end{bash}
\end{mdframed}
It can be imported into the active Python environment using Python's import module: 
\begin{mdframed}[backgroundcolor=mintedbg, linecolor=mintedbg, innerleftmargin=0, innertopmargin=0,innerbottommargin=0]
\begin{python}
    import sciann as sn 
\end{python}
\end{mdframed}
Its mathematical functions are located in the \pythoninline{sn.math} interface. For instance, the function $\textrm{diff}$ is accessed through \pythoninline{sn.math.diff}. The main building blocks of SciANN include: 
\begin{itemize}
\item \pythoninline{sn.Variable}: class to define inputs to the network. 
\item \pythoninline{sn.Field}: class to define outputs of the network. 
\item \pythoninline{sn.Functional}: class to construct a nonlinear neural network approximation. 
\item \pythoninline{sn.Parameter}: class to define a parameter for inversion purposes. 
\item \pythoninline{sn.Data, sn.Tie}: class to define the targets. If there are observations for any variable, the `sn.Data' interface is used when building the optimization model. For physical constraints such as PDEs or equality relations between different variables, the `sn.Tie' interface is designed to build the optimizer.
\item \pythoninline{sn.SciModel}: class to set up the optimization problem, i.e. inputs to the networks, targets (objectives), and the loss function. 
\item \pythoninline{sn.math}: mathematical operations are accessed here. SciANN also support operator overloading, which improves readability when setting up complex mathematical relations such as PDEs.
\end{itemize}

\subsection{An illustrative example: curve fitting}

We illustrate SciANN's capabilities with its application to a curve-fitting problem. Given a set of discrete data, generated from $f(x,y) = \sin(x)\sin(y)$ over the domain $x,y \rightarrow [-\pi, \pi] \times [-\pi, \pi]$, we want to fit a surface, in the form of a neural network, to this dataset. A multi-layer neural network approximating the function~$f$ can be constructed as $\hat{f}: (x, y) \mapsto \mathcal{N}_f(x, y; \mathbf{W}, \mathbf{b})$, with inputs~$x,y$ and output~$\hat{f}$. In the most common mathematical and Pythonic abstraction, the inputs~$x,y$ and output~$\hat{f}$ can be implemented as: 
\begin{mdframed}[backgroundcolor=mintedbg, linecolor=mintedbg, innerleftmargin=0, innertopmargin=0,innerbottommargin=0]
\begin{python}
    x = sn.Variable("x")
    y = sn.Variable("y")
    f = sn.Field("f")
\end{python}
\end{mdframed} 
A 3-layer neural network with 6 neural units and hyperbolic-tangent activation function can then be constructed as 
\begin{mdframed}[backgroundcolor=mintedbg, linecolor=mintedbg, innerleftmargin=0, innertopmargin=0,innerbottommargin=0]
\begin{python}
    f = sn.Functional(
        fields=[f], 
        variables=[x, y], 
        hidden_layers=[6, 6, 6], 
        actf="tanh"
        )
\end{python}
\end{mdframed} 
This definition can be further compressed as 
\begin{mdframed}[backgroundcolor=mintedbg, linecolor=mintedbg, innerleftmargin=0, innertopmargin=0,innerbottommargin=0]
\begin{python}
    f = sn.Functional("f", [x, y], [6, 6, 6], "tanh")
\end{python}
\end{mdframed} 
At this stage, the parameters of the networks, i.e. set of $\mathbf{W}, \mathbf{b}$ for all layers, are randomly initialized. Their current values can be retrieved using the command \pythoninline{get_weights}: 
\begin{mdframed}[backgroundcolor=mintedbg, linecolor=mintedbg, innerleftmargin=0, innertopmargin=0,innerbottommargin=0]
\begin{python}
    f.get_weights()
\end{python}
\end{mdframed} 
One can set the parameters of the network to any desired values using the command \pythoninline{set_weights}.

As another example, a more complex neural network functional as the composition of three blocks, as shown in Fig.~\ref{fig:nn_multinet}, can be constructed as 
\begin{mdframed}[backgroundcolor=mintedbg, linecolor=mintedbg, innerleftmargin=0, innertopmargin=0,innerbottommargin=0]
\begin{python}
    f1 = sn.Functional("f1", [x, y], [4, 4], "tanh")
    f2 = sn.Functional("f2", [x, y], [4, 4], "tanh")
    g = sn.Functional("g", [f1, f2], [4, 4], "tanh")
\end{python}
\end{mdframed} 
Any of these functions can be evaluated immediately or after training using the \pythoninline{eval} function, by providing discrete data for the inputs: 
\begin{mdframed}[backgroundcolor=mintedbg, linecolor=mintedbg, innerleftmargin=0, innertopmargin=0,innerbottommargin=0]
\begin{python}
    f_test = f.eval([x_data, y_data])
    f1_test = f1.eval([x_data, y_data])
    f2_test = f2.eval([x_data, y_data])
    g_test = g.eval([f1_data, f2_data])
\end{python}
\end{mdframed} 

Once the networks are initialized, we set up the optimization problem and \emph{train} the network by minimizing an objective function, i.e. solving the optimization problem for $\mathbf{W}$ and $\mathbf{b}$. The optimization problem for a data-driven curve-fitting is defined as:
\begin{equation}
    \argmin_{\mathbf{W}, \mathbf{b}} \mathcal{L}(\mathbf{W}, \mathbf{b}) := \left\| f(x^*, y^*) - \mathcal{N}_f(x^*, y^*; \mathbf{W}, \mathbf{b}) \right\|,
\end{equation}
where $x^*, y^*$ is the set of all discrete points where $f$ is given. For the loss-function $\left\| \circ \right\|$, we use the mean squared-error norm $\left\| \circ \right\| = \frac{1}{N}\sum_{x^*,y^* \in I} (f(x^*, y^*) - \hat{f}(x^*, y^*))^2$. This problem is set up in SciANN through the \pythoninline{SciModel} class as: 
\begin{mdframed}[backgroundcolor=mintedbg, linecolor=mintedbg, innerleftmargin=0, innertopmargin=0,innerbottommargin=0]
\begin{python}
    m = sn.SciModel(
        inputs = [x, y],
        targets = [f],
        loss_func = "mse",
        optimizer = "adam"
    )
\end{python}
\end{mdframed} 
The \pythoninline{train} model is then used to perform the training and identify the parameters of the neural network:
\begin{mdframed}[backgroundcolor=mintedbg, linecolor=mintedbg, innerleftmargin=0, innertopmargin=0,innerbottommargin=0]
\begin{python}
    m.train([x_data, y_data], [f_data], epochs=400)
\end{python}
\end{mdframed} 

\begin{figure}[H]
    \centering
    \includegraphics[width=1.0\textwidth]{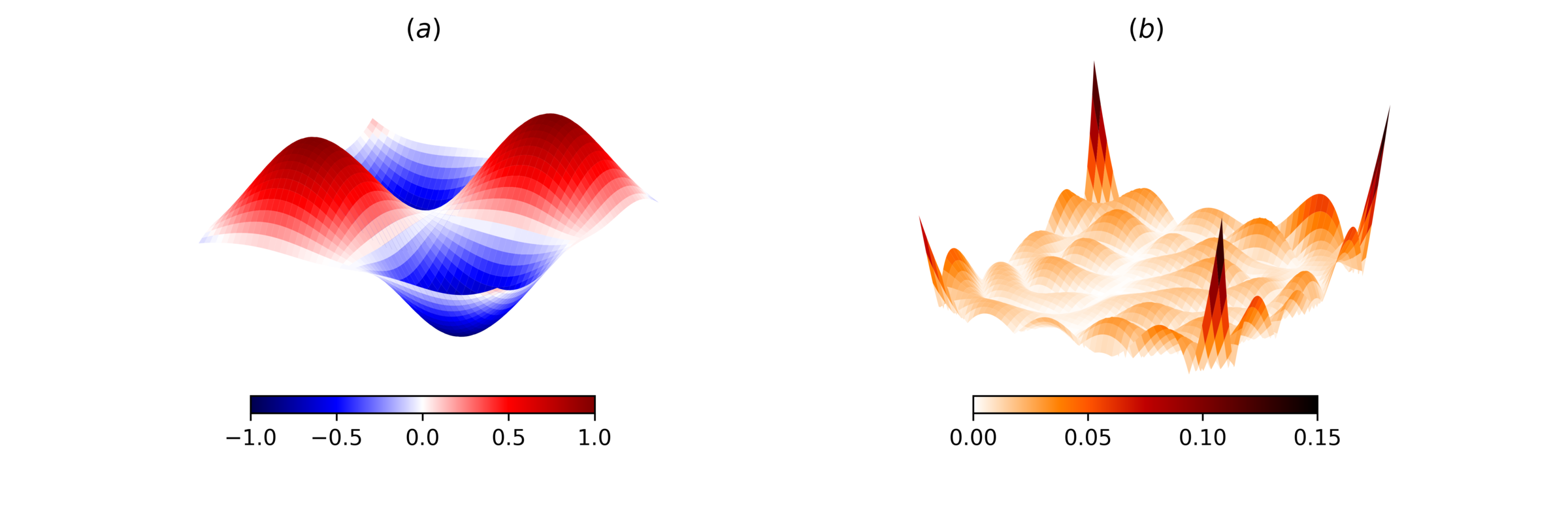}
    \caption{Using SciANN to train a network on synthetic data generated from $\sin(x)\sin(y)$; (a): network predictions; (b): absolute error with respect to true values.}
    \label{fig:sciann-train}
\end{figure}

Once the training is completed, one can set parameters of a \pythoninline{Functional} to be trainable or non-trainable (fixed). For instance, to set $f$ to be non-trainable:
\begin{mdframed}[backgroundcolor=mintedbg, linecolor=mintedbg, innerleftmargin=0, innertopmargin=0,innerbottommargin=0]
\begin{python}
    f1.set_trainable(False)
\end{python}
\end{mdframed} 
The result of this training is shown in Fig.~\ref{fig:sciann-train}, where we have used 400~epochs to perform the training on a dataset generated using a uniform grid of $51 \times 51$.

Since data was generated from $f(x,y) = \sin(x)\sin(y)$, we know that this is a solution to $\Delta f + 2f = 0$, with $\Delta$ as the Laplacian operator. As a first illustration of SciANN for physics-informed deep learning, we can constrain the curve-fitting problem with this `governing equation'. In SciANN, the differentiation operators are evaluated through \pythoninline{sn.math.diff} function. Here, this differential equation can be evaluated as:
\begin{mdframed}[backgroundcolor=mintedbg, linecolor=mintedbg, innerleftmargin=0, innertopmargin=0,innerbottommargin=0]
\begin{python}
    L = diff(fxy,x,order=2) + diff(fxy,y,order=2) + 2*fxy
\end{python}
\end{mdframed} 
with \pythoninline{order} expressing the order of differentiation. 

\begin{figure}[H]
    \centering
    \includegraphics[width=1.0\textwidth]{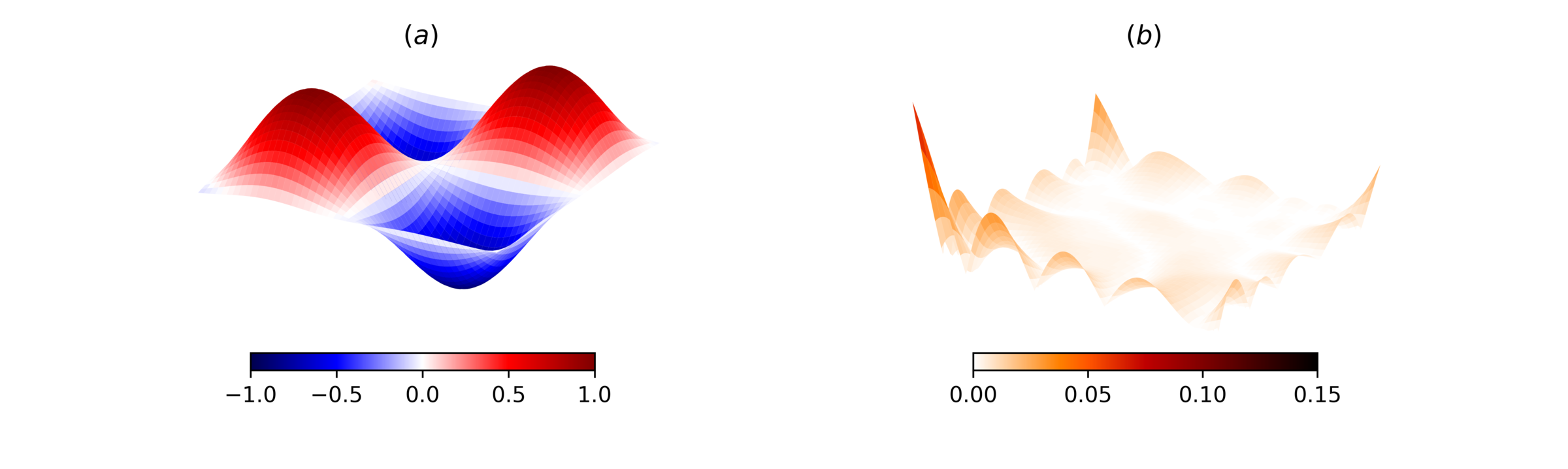}
    \caption{Using SciANN to train a network on synthetic data generated from $\sin(x)\sin(y)$ and imposing the governing equations $f_{,xx} + f_{,yy} - 2f=0$; (a): network predictions; (b): absolute error with respect to true values.}
    \label{fig:sciann-train-L}
\end{figure}

Based on the physics-informed deep learning framework, the governing equation can be imposed through the objective function. The optimization problem can then be defined as 
\begin{equation}
    \argmin_{\mathbf{W}, \mathbf{b}} \mathcal{L}(\mathbf{W}, \mathbf{b}) := \left\| f(x^*, y^*) - \hat{f}(x^*, y^*) \right\| + \left\| \Delta \hat{f}(x^*, y^*) + 2 \hat{f}(x^*, y^*)\right\|,
\end{equation}
and implemented in SciANN as 
\begin{mdframed}[backgroundcolor=mintedbg, linecolor=mintedbg, innerleftmargin=0, innertopmargin=0,innerbottommargin=0]
\begin{python}
    m = SciModel([x, y], [fxy, L])
    m.train([x_mesh, y_mesh], [(ids_data, fxy_data), 'zero'], epochs=400)
\end{python}
\end{mdframed} 
Note that while the inputs are the same as for the previous case, the optimization model is defined with two targets, \pythoninline{fxy} and \pythoninline{L}. The training data for \pythoninline{fxy} remains the same; the sampling grid, however, can be expanded further as `physics' can be imposed everywhere. A sampling grid $101 \times 101$ is used here, where data is only given at the same locations as the previous case, i.e. on the $51 \times 51$ grid. To impose target \pythoninline{L}, it is simply set to \pythoninline{'zero'}. The new result is shown in Fig.~\ref{fig:sciann-train-L}. We find that, for the same network size and training parameters, incorpo rating the `physics' reduces the error significantly.


Once the training is completed, the weights $\mathbf{W}, \mathbf{b}$ for all layers can be saved using the command \pythoninline{save_weights}, for future use. These weights can be later used to initialize a network of the same structure using \pythoninline{load_weights_from} keyword in \pythoninline{SciModel}.

\section{Application of SciANN to Physics-Informed Deep Learning}
In this section, we explore how to use SciANN to solve and discover some representative case studies of physics-informed deep learning. 

\subsection{Burgers equation}
As the first example, we illustrate the use of SciANN to solve the Burgers equation, which arises in fluid mechanics, acoustics, and traffic flow~\cite{dafermos-claws}. Following~\cite{Raissi2018}, we explore the governing equation:
\begin{equation}\label{eq:burgers1}
    u_{,t} + u u_{,x} - (0.01 / \pi) u_{,xx} = 0, \quad t \in [0, 1], \quad x \in [-1,1],
\end{equation}
subject to initial and boundary conditions $u(t=0, x) = -\sin(\pi x)$ and $u(t, x=\pm 1) = 0$, respectively. The solution variable~$u$ can be approximated by~$\hat{u}$, defined in the form of a nonlinear neural network as $\hat{u}: (t,x) \mapsto \mathcal{N}_u(t,x; \mathbf{W, \mathbf{b}})$. The network used in~\cite{Raissi2018} consists of 8~hidden layers, each with 20~neurons, and with $\tanh$ activation function, and can be defined in SciANN as:
\begin{mdframed}[backgroundcolor=mintedbg, linecolor=mintedbg, innerleftmargin=0, innertopmargin=0,innerbottommargin=0]
\begin{python}
    t = sn.Variable("t")
    x = sn.Variable("x")
    u = sn.Functional("u", [t, x], 8*[20], "tanh")
\end{python}
\end{mdframed} 

To set up the optimization problem, we need to identify the targets. The first target, as used in the PINN framework, is the PDE in Eq.~\eqref{eq:burgers1}, and is defined in SciANN as: 
\begin{mdframed}[backgroundcolor=mintedbg, linecolor=mintedbg, innerleftmargin=0, innertopmargin=0,innerbottommargin=0]
\begin{python}
    import sciann.math.diff as diff
    L1 = diff(u, t) + u*diff(u, x) - (0.01/pi)*diff(u, x, order=2)
\end{python}
\end{mdframed} 
To impose boundary conditions, one can define them as continuous mathematical functions defined at all sampling points:
\begin{equation}\label{eq:burgers2}
\begin{split}
    L_2 &:= (1-\mathrm{sign}(t - t_{\textrm{min}}))(u + sin(\pi x)), \\
    L_3 &:= (1 - \mathrm{sign}(x - x_{\textrm{min}}))u, \\
    L_4 &:= (1 + \mathrm{sign}(x - x_{\textrm{max}}))u,
\end{split}
\end{equation}
For instance, $L_2$~is zero at all sampling points except for $t<t_{\textrm{min}}$, which is chosen as $t_0 + \mathrm{tol}$ . Instead of~$\mathrm{sign}$, one can use smoother functions such as~$\tanh$. In this way, the optimization model can be set up as: 
\begin{mdframed}[backgroundcolor=mintedbg, linecolor=mintedbg, innerleftmargin=0, innertopmargin=0,innerbottommargin=0]
\begin{python}
    m = sn.SciModel([t, x], [L1, L2, L3, L4], "mse", "Adam")
\end{python}
\end{mdframed} 
In this case, all targets should `vanish', therefore the training is done as:
\begin{mdframed}[backgroundcolor=mintedbg, linecolor=mintedbg, innerleftmargin=0, innertopmargin=0,innerbottommargin=0]
\begin{python}
    m.train(
        [x_data, t_data], 
        ['zeros', 'zeros', 'zeros', 'zeros'], 
        batch_size=256, epochs=10000
    )
\end{python}
\end{mdframed} 

An alternative approach to define the boundary conditions in SciANN is to define the target in the \pythoninline{sn.SciModel} as the variable of interest and pass the `ids' of training data where the conditions should be imposed. This is achieved as: 
\begin{mdframed}[backgroundcolor=mintedbg, linecolor=mintedbg, innerleftmargin=0, innertopmargin=0,innerbottommargin=0]
\begin{python}
    m = sn.SciModel([t, x], [L1, u], "mse", "Adam")
    m.train(
        [x_data, t_data], 
        ['zeros',  (ids_ic_bc, U_ic_bc)], 
        batch_size=256, epochs=10000
    )
\end{python}
\end{mdframed} 
Here, \pythoninline{ids_ic_bc} are ids associated with collocation points \pythoninline{(t_data, x_data)} where the initial condition and boundary condition are given. An important point to keep in mind is that if the number of sampling points where boundary conditions are imposed is a very small portion, the mini-batch optimization parameter \pythoninline{batch_size} should be set to a large number to guarantee consistent mini-batch optimization. Otherwise, some mini-batches may not acquire any data on the boundary and therefore not generate the correct gradient for the gradient-descent update. Also worth noting is that setting governing relations to `zero' is conveniently done in SciANN. 

\begin{figure}[H]
    \centering
    \includegraphics[width=1.0\textwidth]{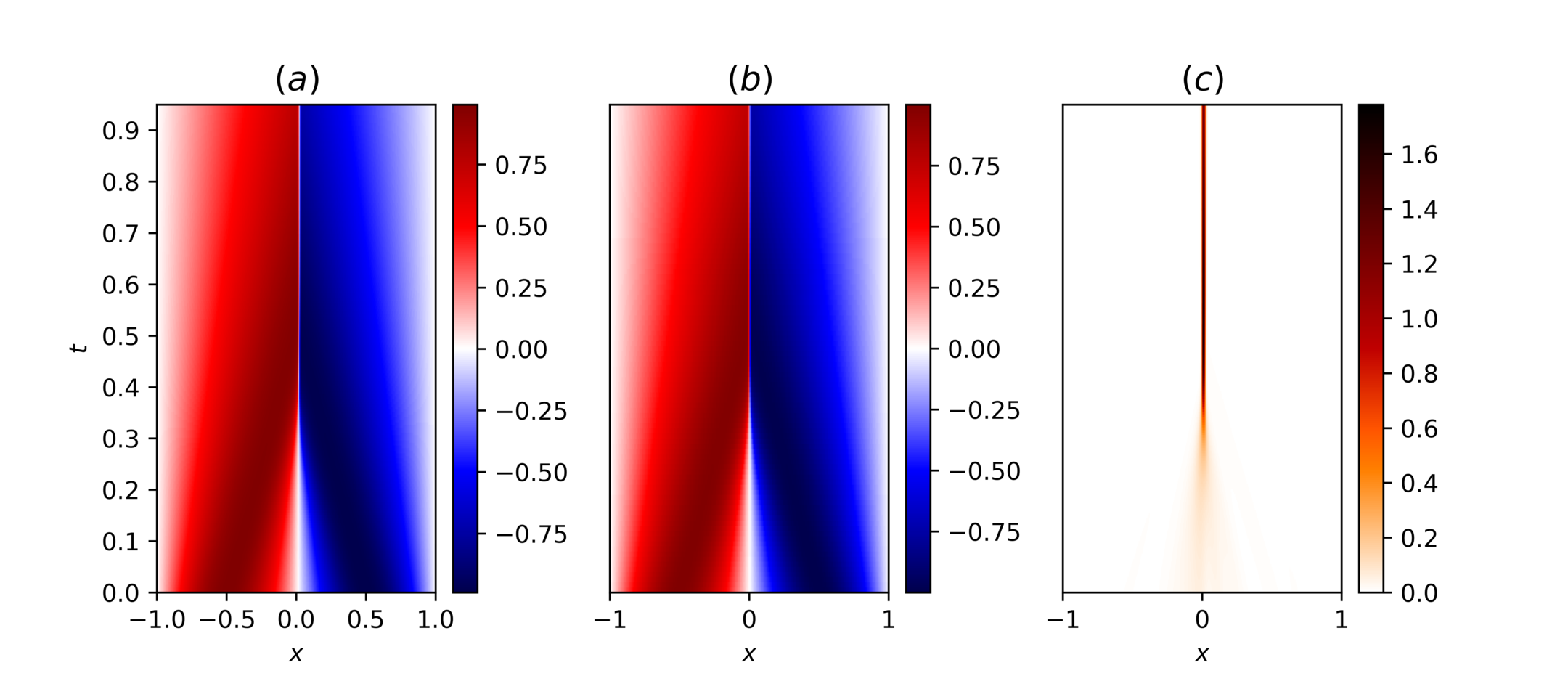}
    \caption{Solution of the Burgers equation using PINN. (a)~True solution for~$u$; (b)~PINN predicted values~$\hat{u}$; (c)~Absolute error between true and predicted values, $|u-\hat{u}|$.}
    \label{fig:burgers}
\end{figure}

The result of solving the Burgers equation using the deep learning framework is shown in Fig.~\ref{fig:burgers}. The results match the exact solution accurately, and reproduce the formation of a shock (self-sharpening discontinuity) in the solution at $x=0$.

\subsection{Data driven discovery of Navier--Stokes equations}

As a second example, we show how SciANN can be used for discovery of partial differential equations. We choose the incompressible Navier--Stokes problem used in~\cite{Raissi2019}. The equations are: 
\begin{equation}\label{eq:navier1}
\begin{split}
    u_{,t} + p_{,x} + \lambda_1(u u_{,x} + v u_{,y}) - \lambda_2 (u_{,xx} + u_{,yy}) &= 0, \\
    v_{,t} + p_{,y} + \lambda_1(u v_{,x} + v v_{,y}) - \lambda_2 (v_{,xx} + v_{,yy}) &= 0, \\
\end{split}
\end{equation}
where~$u$ and~$v$ are components of velocity field in~$x$ and~$y$ directions, respectively, $p$~is the density-normalized pressure, $\lambda_1$ should be identically equal to~1 for Newtonian fluids, and~$\lambda_2$ is the kinematic viscosity. The true value of the parameters to be identified are $\lambda_1=1$ and $\lambda_2=0.01$. Given the assumption of fluid incompressibility, we use the divergence-free form of the equations, from which the components of the velocity are obtained as: 
\begin{equation}\label{eq:navier2}
    u = \psi_{,y}, \quad v = -\psi_{,x},
\end{equation}
where $\psi$~is the potential function. 

Here, the independent field variables~$p$ and~$\psi$ are approximated as $p(t,x,y) \approx \hat{p}(t,x,y)$ and $\psi(t,x,y) \approx \hat{\psi}(t,x,y)$, respectively, using nonlinear artificial neural networks as $\hat{p}: (t,x,y) \mapsto \mathcal{N}_p(t,x,y; \mathbf{W}, \mathbf{b})$ and $\hat{\psi}: (t,x,y) \mapsto \mathcal{N}_{\psi}(t,x,y; \mathbf{W}, \mathbf{b})$. Using the same network size and activation function that was used in~\cite{Raissi2019}, we set up the neural networks in SciANN as:
\begin{mdframed}[backgroundcolor=mintedbg, linecolor=mintedbg, innerleftmargin=0, innertopmargin=0,innerbottommargin=0]
\begin{python}
    p = sn.Functional("p", [t, x, y], 8*[20], 'tanh')
    psi = sn.Functional("psi", [t, x, y], 8*[20], 'tanh')
\end{python}
\end{mdframed} 
Note that this way of defining the networks results in \emph{two} separate networks for~$p$ and~$\psi$, which we find more suitable for many problems. To replicate the one-network model used in the original study, one can use: 
\begin{mdframed}[backgroundcolor=mintedbg, linecolor=mintedbg, innerleftmargin=0, innertopmargin=0,innerbottommargin=0]
\begin{python}
    p, psi = sn.Functional(["p", "psi"], [t, x, y], 8*[20], 'tanh').split()
\end{python}
\end{mdframed} 

\begin{figure}[H]
    \centering
    \includegraphics[width=1\textwidth]{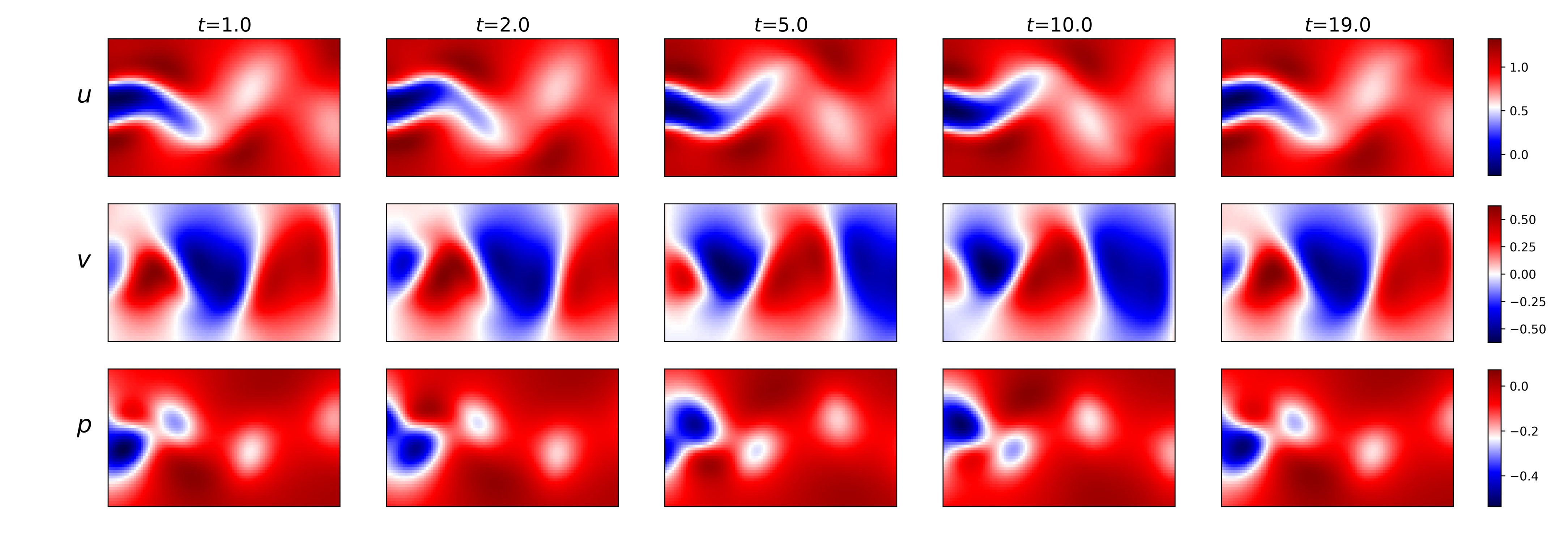}
    \caption{Predicted values from the PINN framework, for the field variables~$u$, $v$ and~$p$, at different times~$t$. The parameters are identified as $\lambda_1=0.9967$ and $\lambda_2=0.0110$. The predictions are the same as those reported in~\cite{Raissi2019}.}
    \label{fig:navier-pinn}
\end{figure}

Here, the objective is to identify parameters $\lambda_1$ and $\lambda_2$ of the Navier--Stokes equations~\eqref{eq:navier1} on a dataset with given velocity field. Therefore, we need to define these as trainable parameters of the network. This is done using \pythoninline{sn.Parameter} interface as:
\begin{mdframed}[backgroundcolor=mintedbg, linecolor=mintedbg, innerleftmargin=0, innertopmargin=0,innerbottommargin=0]
\begin{python}
    lamb1 = sn.Parameter(0.0, [x, y], name="lamb1")
    lamb2 = sn.Parameter(0.0, [x, y], name="lamb2")
\end{python}
\end{mdframed} 
Note that these parameters are initialized with a value of~$0.0$. The required derivatives in Equations~\eqref{eq:navier1} and~\eqref{eq:navier2} are evaluated as: 
\begin{mdframed}[backgroundcolor=mintedbg, linecolor=mintedbg, innerleftmargin=0, innertopmargin=0,innerbottommargin=0]
\begin{python}
    u, v = diff(psi,y), -diff(psi,x)
    u_t, v_t = diff(u,t), diff(v,t)
    u_x, u_y = diff(u,x), diff(u,y)
    v_x, v_y = diff(v,x), diff(v,y)
    u_xx, u_yy = diff(u,x,order=2), diff(u,y,order=2)
    v_xx, v_yy = diff(v,x,order=2), diff(v,y,order=2)
    p_x, p_y = diff(p,x), diff(p,y)
\end{python}
\end{mdframed} 
with `order' indicating the order of differentiation. We can now set up the targets of the problem as:
\begin{mdframed}[backgroundcolor=mintedbg, linecolor=mintedbg, innerleftmargin=0, innertopmargin=0,innerbottommargin=0]
\begin{python}
    L1 = u_t + p_x + lamb1*(u*u_x + v*u_v) - lamb2*(u_xx + u_yy)
    L2 = v_t + p_y + lamb1*(u*v_x + v*v_y) - lamb2*(v_xx + v_yy)
    L3 = u
    L4 = v
\end{python}
\end{mdframed} 
The optimization model is now set up as:
\begin{mdframed}[backgroundcolor=mintedbg, linecolor=mintedbg, innerleftmargin=0, innertopmargin=0,innerbottommargin=0]
\begin{python}
    m = sn.SciModel([t, x, y], [L1, L2, L3, L4], "mse", "Adam")
    m.train([t_data, x_data, y_data],
            ['zeros', 'zeros', u_data, v_data],
            batch_size=64, epochs=10000)
\end{python}
\end{mdframed} 
where only training points for~$u$ and~$v$ are provided, as in~\cite{Raissi2019}. The results are shown in Fig.~\ref{fig:navier-pinn}. 

\subsection{Discovery of nonlinear solid mechanics with von~Mises plasticity}

Here, we illustrate the use of PINN for solution and discovery of nonlinear solid mechanics. We use the von~Mises elastoplastic constitutive model, which is commonly used to describe mechanical behavior of solid materials, in particular metals. Elastoplasticity relations give rise to inequality constraints on the governing equations~\cite{simohughes-ci}, and, therefore, compared to the Navier--Stokes equations, they pose a different challenge to be incorporated in PINN. The elastoplastic relations for a plane-strain problem are: 
\begin{equation}\label{eq:vonmises1}
\begin{split}
    \sigma_{ij, j} + f_i &= 0, \\
    \sigma_{ij} &= s_{ij} - p\delta_{ij},  \\
    p &= -\sigma_{kk}/3 = -(\lambda + 2/3\mu) \varepsilon_v, \\
    s_{ij} &= 2\mu e^e_{ij}, \\
    \varepsilon_{ij} &= (u_{i,j} + u_{j,i})/2 = e_{ij} + \varepsilon_v\delta_{ij}/3, \\
    \varepsilon_v &= \varepsilon_{kk} = \varepsilon_{xx} + \varepsilon_{yy}, \\
    e_{ij} &= e^e_{ij} + e^p_{ij}. \\
     \\
\end{split}
\end{equation}
Here, the summation notation is used with $i,j,k \in \{x, y\}$. $\sigma_{ij}$ are components of the Cauchy stress tensor, and~$s_{ij}$ and~$p$ are its deviatoric components and its pressure invariant, respectively. $\varepsilon_{ij}$ are components of the infinitesimal strain tensor derived from the displacements $u_x, u_y$, and $e_{ij}$ and $\varepsilon_v$ are its deviatoric and volumetric components, respectively.

According to the von~Mises plasticity model, the admissible state of stress is defined inside the cylindrical yield surface $\mathcal{F} = \mathcal{F}(\sigma_{ij})$ as $\mathcal{F} := q - \sigma_Y \le 0$. Here, $q$ is the equivalent stress defined as $q=\sqrt{3/2 s_{ij} s_{ij}}$. Assuming the associative flow rule, the plastic strain components are: 
\begin{equation}\label{eq:vonmises2}
    \varepsilon^p_{ij} \equiv e^p_{ij} = \bar{e}^p \frac{\partial \mathcal{F}}{\partial \sigma_{ij}} =  \bar{\varepsilon}^p \frac{3}{2}\frac{s_{ij}}{q},
\end{equation}
where $\bar{\varepsilon}^p$ is the equivalent plastic strain, subject to $\bar{\varepsilon}^p \ge 0$. For the von~Mises model, it can be shown that $\bar{\varepsilon}^p$ is evaluated as 
\begin{equation}\label{eq:vonmises3}
    \bar{\varepsilon}^p = \bar{\varepsilon}  - \frac{\sigma_Y}{3\mu} \ge 0,
\end{equation}
where $\bar{\varepsilon}$ is the total equivalent strain, defined as $\bar{\varepsilon}=\sqrt{2/3e_{ij}e_{ij}}$.
Note that for von~Mises plasticity, the volumetric part of plastic strain tensor is zero, $\varepsilon_v^p=0$. Finally, the parameters of this model include the Lam\'e elastic parameters~$\lambda$ and~$\mu$, and the yield stress~$\sigma_Y$. 

We use a classic example to illustrate our framework: a perforated strip subjected to uniaxial extension \cite{zienkiewicz1969elasto, simohughes-ci}. Consider a plate of dimensions $200~\text{mm}\times360~\text{mm}$, with a circular hole of diameter $100~\text{mm}$ located in the center of the plate. The plate is subjected to extension displacements of $\delta=1~\text{mm}$ along the short edge, under plane-strain condition, and without body forces, $f_i=0$. The parameters are $\lambda=19.44~\textrm{GPa}$, $\mu=29.17~\textrm{GPa}$ and $\sigma_Y=243.0~\textrm{MPa}$. Due to symmetry, only a quarter of the domain needs to be considered in the simulation. The synthetic data is generated from a high-fidelity FEM simulation using COMSOL software~\cite{COMSOL} on a mesh of approximately 13,000 quartic triangular elements. The plate undergoes significant plastic deformation around the circular hole. This results in localized deformation in the form of a shear band. While the strain exhibits localization, the stress field remains continuous and smooth---a behavior that is due to the choice of a perfect-plasticity model with no hardening.

\begin{figure}[H]
    \centering
    \includegraphics[width=0.8\textwidth]{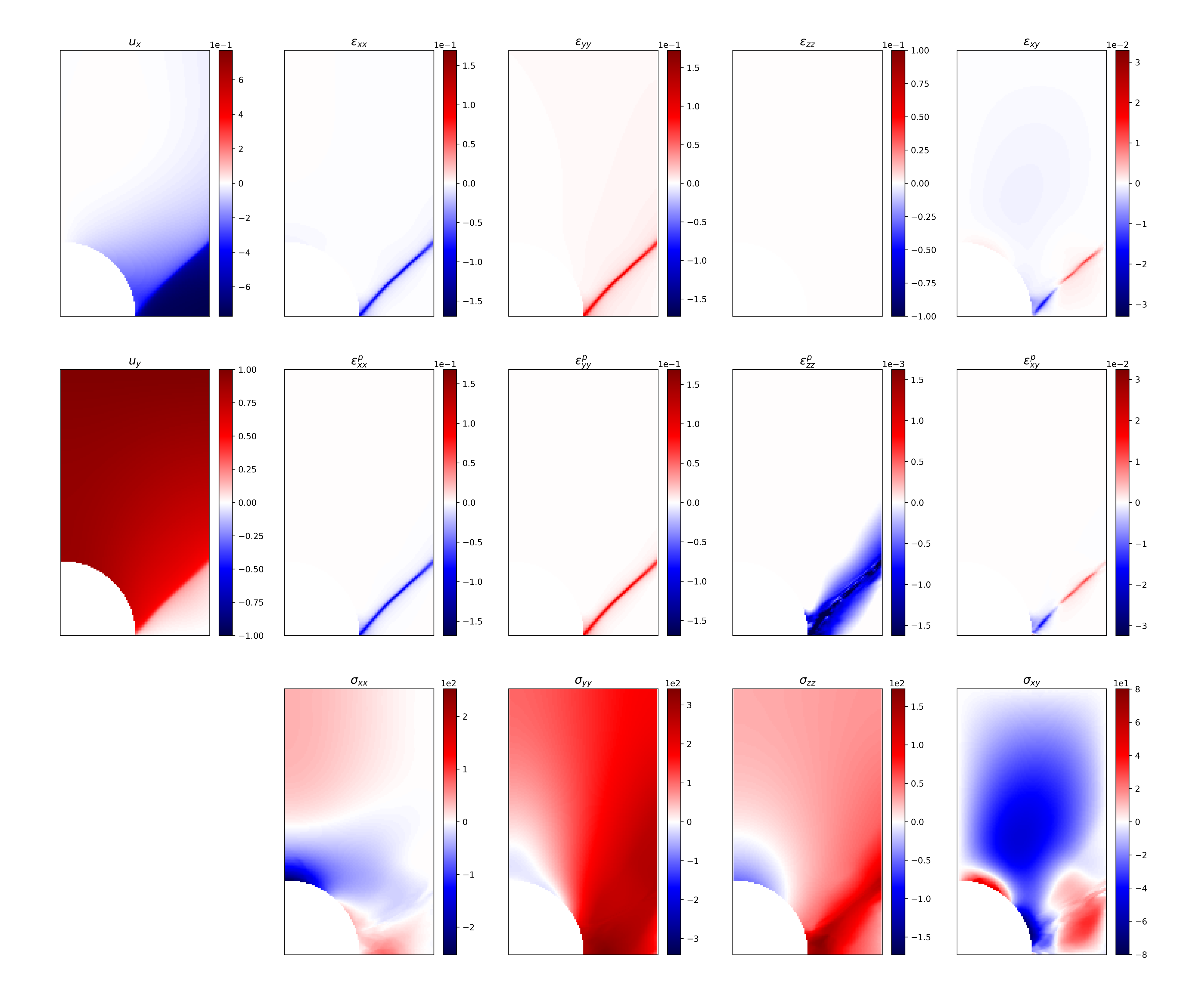}
    \caption{The predicted values from the PINN framework for displacements, strains, plastic strains and stresses. The inverted parameters are $\lambda=18.3~\textrm{GPa}$, $\mu=27.6~\textrm{GPa}$ and $\sigma_Y=243.0~\textrm{MPa}$. }
    \label{fig:vonmises2}
\end{figure}

Following the approach proposed in~\cite{Haghighat2020}, we approximate displacement and stress components $u_x, u_y, \sigma_{xx}, \sigma_{yy}, \sigma_{zz}, \sigma_{xy}$ with nonlinear neural networks as:
\begin{equation}\label{eq:vonmises4}
\begin{split}
    \hat{u}_x&: (x, y) \mapsto \mathcal{N}_{u_x}(x, y; \mathbf{W}, \mathbf{b}) \\
    \hat{v}_x&: (x, y) \mapsto \mathcal{N}_{v_x}(x, y; \mathbf{W}, \mathbf{b}) \\
    \hat{\sigma}_{xx}&: (x, y) \mapsto \mathcal{N}_{\sigma_{xx}}(x, y; \mathbf{W}, \mathbf{b}) \\
    \hat{\sigma}_{yy}&: (x, y) \mapsto \mathcal{N}_{\sigma_{yy}}(x, y; \mathbf{W}, \mathbf{b}) \\
    \hat{\sigma}_{zz}&: (x, y) \mapsto \mathcal{N}_{\sigma_{zz}}(x, y; \mathbf{W}, \mathbf{b})\\
    \hat{\sigma}_{xy}&: (x, y) \mapsto \mathcal{N}_{\sigma_{xy}}(x, y; \mathbf{W}, \mathbf{b})
\end{split}
\end{equation}
Note that due to plastic deformation, the out-of-plane stress $\sigma_{zz}$ is not predefined, and therefore we also approximate it with a neural network. These neural networks and parameters $\lambda, \mu, \sigma_Y$ are defined as follows:
\begin{mdframed}[backgroundcolor=mintedbg, linecolor=mintedbg, innerleftmargin=0, innertopmargin=0,innerbottommargin=0]
\begin{python}
    ux = sn.Functional('ux', [x, y], 4*[50], 'tanh')
    uy = sn.Functional('uy', [x, y], 4*[50], 'tanh')
    sxx = sn.Functional('sxx', [x, y], 4*[50], 'tanh')
    syy = sn.Functional('syy', [x, y], 4*[50], 'tanh')
    szz = sn.Functional('szz', [x, y], 4*[50], 'tanh')
    sxy = sn.Functional('sxy', [x, y], 4*[50], 'tanh')
    lmbd = sn.Paramater(1.0, [x, y])
    mu = sn.Paramater(1.0, [x, y])
    sy = sn.Paramater(1.0, [x, y])
\end{python}
\end{mdframed} 
The kinematic relations, deviatoric stress components and plastic strains can be defined as:
\begin{mdframed}[backgroundcolor=mintedbg, linecolor=mintedbg, innerleftmargin=0, innertopmargin=0,innerbottommargin=0]
\begin{python}
    # Total strain components
    Exx = diff(ux, x)  
    Eyy = diff(uy, y)
    Exy = (diff(ux, y) + diff(uy, x))/2
    Evol = Exx + Eyy
    # Deviatoric strain components 
    exx = Exx - Evol/3
    eyy = Eyy - Evol/3
    ezz = - Evol/3 # Ezz=0 (plane strain)
    ebar = sn.math.sqrt(2/3*(exx**2 + eyy**2 + ezz**2 + 2*exy**2))
    # Deviatoric stress components 
    prs = -(sxx + syy + szz)/3
    dxx = sxx + prs 
    dyy = syy + prs 
    dzz = szz + prs 
    q = sn.math.sqrt(3/2*(dxx**2 + dyy**2 + dzz**2 + 2*sxy**2))
    # Plastic strain components 
    pebar = sn.math.relu(ebar - sy/(3*mu))
    pexx = 1.5 * pebar * sxx / q
    peyy = 1.5 * pebar * syy / q
    pezz = 1.5 * pebar * szz / q
    pexy = 1.5 * pebar * sxy / q
    # Yield surface 
    F = q - sy
\end{python}
\end{mdframed} 
The operator-overloading abstraction of SciANN improves readability significantly. Assuming access to the measured data for variables $u_x$, $u_y$, $\sigma_{xx}$, $\sigma_{yy}$, $\sigma_{zz}$, $\sigma_{xy}$, $\varepsilon_{xx}$, $\varepsilon_{yy}$, $\varepsilon_{xy}$, the optimization targets for training data can be described using the \pythoninline{L* = sn.Data(*)}, where $\mathrm{*}$ refers to each variable. The physics-informed constraints are set as: 
\begin{mdframed}[backgroundcolor=mintedbg, linecolor=mintedbg, innerleftmargin=0, innertopmargin=0,innerbottommargin=0]
\begin{python}
    # Volumetric (hydro-static) stress 
    L1 = sn.Tie(prs, -kappa*Evol)
    # Stress relations 
    L2 = sn.Tie(dxx, 2*mu*(exx - pexx))
    L3 = sn.Tie(dyy, 2*mu*(eyy - peyy))
    L4 = sn.Tie(dzz, 2*mu*(ezz - pezz))
    L5 = sn.Tie(dxy, 2*mu*(exy - pexy))
    # Yield surface 
    # Penalize the positive part 
    L6 = sn.math.relu(F)
    # Momentum relations
    L7 = sn.diff(sxx, x) + sn.diff(sxy, y)
    L8 = sn.diff(sxy, x) + sn.diff(syy, y)
\end{python}
\end{mdframed} 
We use 2,000 data points from this reference solution, randomly distributed in the simulation domain, to provide the training data. The PINN training is performed using networks with 4~layers, each with 100~neurons, and with a hyperbolic-tangent activation function. The optimization parameters are the same as those used in~\cite{Haghighat2020}. The results predicted by the PINN approach match the reference results very closely, as evidenced by: (1)~the very small errors in each of the components of the solution, except for the out-of-plane plastic strain components (Fig.~\ref{fig:vonmises2}); and (2)~the precise identification of yield stress $\sigma_Y$ and relatively accurate identification of elastic parameters~$\lambda$ and~$\mu$, yielding estimated values $\lambda=18.3~\textrm{GPa}$, $\mu=27.6~\textrm{GPa}$ and $\sigma_Y=243.0~\textrm{MPa}$. %

\section{Application to Variational PINN}
Neural networks have recently been used to solve the \emph{variational form} of differential equations as well~\cite{Weinan2018, Berg2018}. In a recent study~\cite{Kharazmi2020}, the vPINN framework for solving PDEs was introduced and analyzed. Like PINN, it is based on graph-based automatic differentiation. The authors of~\cite{Kharazmi2020} suggest a Petrov--Galerkin approach, where the test functions are chosen differently from the trial functions. For the test functions, they propose the use of polynomials that vanish on the boundary of the domain. Here, we illustrate how to use SciANN for vPINN, and we show how to construct proper test functions using neural networks. 

Consider the steady-state heat equation subject to Dirichlet boundary conditions and a known heat source~$f(x,y)$~\cite{Kharazmi2020}: 
\begin{equation}\label{eq:vpinn1}
\Delta T + f(x,y) = 0, \quad x,y\in[-1,1]\times[-1,1], 
\end{equation}
subject to the following boundary conditions: 
\begin{equation}\label{eq:vpinn2}
\begin{split}
T(x=\pm 1, y) &= \sin(2\pi y), \\
T(x, y=\pm 1) &= 0,
\end{split}
\end{equation}
and a heat source:
\begin{equation}\label{eq:vpinn3}
\begin{split}
f(x,y) &= \sin\left(2\pi y\right)\left(20\tanh\left(10x\right)\left(10{\tanh\left(10x\right)}^2-10\right)-\frac{2\pi ^2\sin\left(2\pi x\right)}{5}\right) \\
&~~ -4\pi ^2\sin\left(2\pi y\right)\left(\tanh\left(10x\right)+\frac{\sin\left(2\pi x\right)}{10}\right).
\end{split}
\end{equation}
The analytical solution to this problem is:
\begin{equation}\label{eq:vpinn4}
    T(x,y) = \left(0.1\sin(2\pi x) + \tanh(10 x)\right) \sin(2\pi y).
\end{equation}

The weak form of Eq.~\eqref{eq:vpinn1} is expressed as: 
\begin{equation}\label{eq:vpinn5}
    \int_{\Omega} \left[ \nabla Q \cdot \nabla T + Q \cdot f(x,y) \right]\,dV = \int_{\partial \Omega} Q q_n\,dS, 
\end{equation}
where $\Omega$~is the domain of the problem, $\partial\Omega$~is the boundary of the domain, $q_n$~is the boundary heat flux, and $Q$~is the test function. The trial space for the temperature field $T$ is constructed by a neural network as $T: (x,y) \mapsto \mathcal{N}_T(x, y; \mathbf{W}, \mathbf{b})$. For the test space~$Q$, the authors of~\cite{Kharazmi2020} suggest the use of polynomials that satisfy the boundary conditions. However, considering the universal approximation capabilities of the neural networks, we suggest that this step is unnecessary, and a general neural network can be used as the test function. Note that test functions should satisfy continuity requirements as well as boundary conditions. A multi-layer neural network with any nonlinear activation function is a good candidate for the continuity requirements. To satisfy the boundary conditions, we can simply train the test functions to vanish on the boundary. Note that this step is associated to the construction of proper test function and is done as a preprocessing step. Once the test functions satisfy the (homogeneous) boundary conditions, there is no need to further train them, and therefore their parameters can be set to non-trainable at this stage. We also find that there is no need for the $\mathcal{N}_T$ and $\mathcal{N}_Q$ networks to be of the same size, or use the same activation functions. 

Therefore, the test function~$Q$ is defined as $Q: (x,y) \mapsto \mathcal{N}_{Q}(x,y; \bar{\mathbf{W}}, \bar{\mathbf{b}})$ subject to $Q(x=\pm 1, y) = Q(x, y\pm=1) = 0$. Here, overbar weights and biases $\bar{\mathbf{W}}, \bar{\mathbf{b}}$ indicate that their values are predefined and fixed (non-trainable). Therefore, the boundary flux integral on the right side of Eq.~\eqref{eq:vpinn5} vanishes, and the resulting week form can be expressed as  
\begin{equation}\label{eq:vpinn6}
    \int_{\Omega} \nabla Q \cdot \nabla T + Q \cdot f(x,y)\,dV = 0.
\end{equation}
The problem can be defined in SciANN as follows. The first step is to construct proper test function: 
\begin{mdframed}[backgroundcolor=mintedbg, linecolor=mintedbg, innerleftmargin=0, innertopmargin=0,innerbottommargin=0]
\begin{python}
    Q = sn.Functional('Q', [x, y], 4*[20], 'sigmoid')
    m = sn.SciModel([x, y], [Q])
    m.train([x_data, y_data], [Q_data])
    Q.set_trainable(False)
\end{python}
\end{mdframed} 
As discussed earlier, \pythoninline{Q_data} takes a value of~$0.0$ for training points on the boundary and random values at interior quadrature points. Additionally, parameters of~$Q$ are set to non-trainable at the end of this step. The trial function~$T$ and the target weak form in Eq.~\eqref{eq:vpinn6} are now implemented as:
\begin{mdframed}[backgroundcolor=mintedbg, linecolor=mintedbg, innerleftmargin=0, innertopmargin=0,innerbottommargin=0]
\begin{python}
    T = sn.Functional('T', [x, y], 4*[20], 'tanh')
    Q_x, Q_y = diff(Q, x), diff(Q, y)
    T_x, T_y = diff(T, x), diff(T, y)
    # New variables for body force and volume information
    fxy = sn.Variable('fxy')
    vol = sn.Variable('vol')
    # The variational target
    J = (Q_x*T_x + Q_y*T_y + Q*fxy) * vol 
\end{python}
\end{mdframed} 
Since the variational relation~\eqref{eq:vpinn6} takes an integral form, we need to perform a domain integral. Therefore, the volume information should be passed to the network along with the body-force information at the quadrature points. This is achieved by introducing two new SciANN variables as the inputs to the network. The optimization model is then defined as:
\begin{mdframed}[backgroundcolor=mintedbg, linecolor=mintedbg, innerleftmargin=0, innertopmargin=0,innerbottommargin=0]
\begin{python}
    m = sn.SciModel([x, y, vol], [J, T], "mse")
    m.train(
        [x_data, y_data, vol_data, fxy_data], 
        ['zeros', (bc_ids, bc_vals)],
    )
\end{python}
\end{mdframed} 
The second target on~$T$ imposes the boundary conditions at specific quadrature points \pythoninline{bc_ids}. 

\begin{figure}[H]
    \centering
    \includegraphics[width=1.0\textwidth]{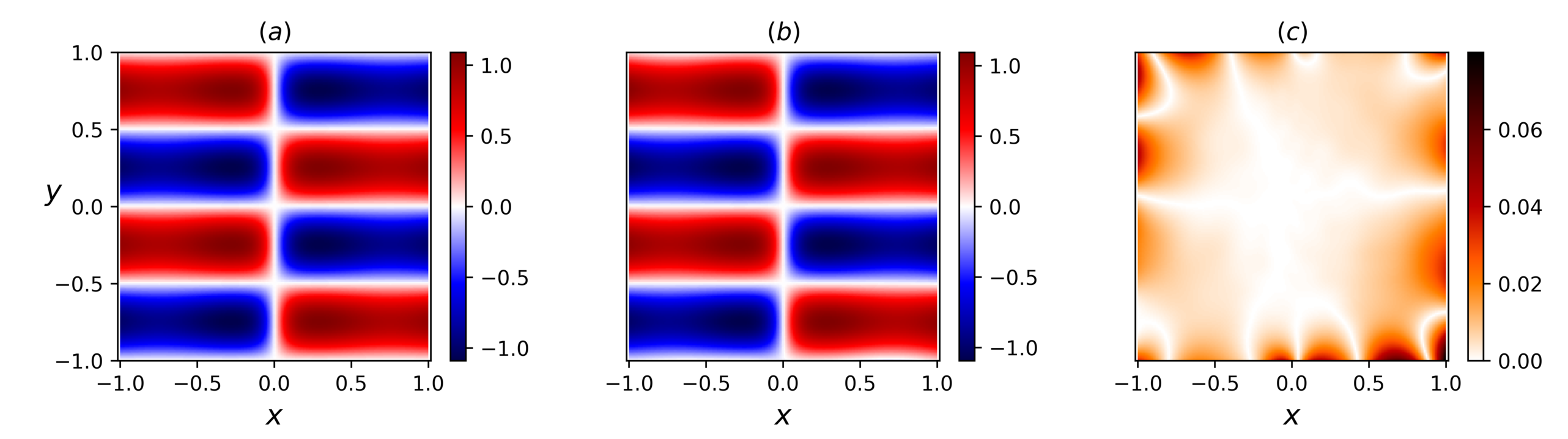}
    \caption{Solution of a steady-state heat equation using the vPINN framework. (a)~True temperature field, $T(x,y)$. (b)~Temperature field predicted by the neural network, $\hat{T}(x,y)$. (c)~Absolute error between true and predicted values, $|T(x,y)-\hat{T}(x,y)|$. }
    \label{fig:vpinn}
\end{figure}

Following the details in~\cite{Kharazmi2020}, we perform the integration on a $70\times 70$~grid. The results are shown in Fig.~\ref{fig:vpinn}, which are very similar to those reported in~\cite{Kharazmi2020}.

\section{Conclusions}

In this paper, we have introduced the open-source deep-learning package, SciANN, designed specifically to facilitate physics-informed simulation, inversion, and discovery in the context of computational science and engineering problems. It can be used for regression and physics-informed deep learning with minimal effort on the neural network setup. It is based on Tensorflow and Keras packages, and therefore it inherits all the high-performance computing capabilities of Tensorflow back-end, including CPU/GPU parallelization capabilities. 

The objective of this paper is to introduce an environment based on a modern implementation of graph-based neural network and automatic differentiation, to be used as a platform for scientific computations. In a series of examples, we have shown how to use SciANN for curve-fitting, solving PDEs in strong and weak form, and for model inversion in the context of physics-informed deep learning. The examples presented here as well as the package itself are all open-source, and available in the github repository \href{https://github.com/sciann}{github.com/sciann}.

\section*{Acknowledgments}
This work was funded by the KFUPM-MIT collaborative agreement `Multiscale Reservoir Science'.

\section*{References}

\bibliographystyle{elsarticle-num}
\bibliography{ml_references}

\begin{thebibliography}{10}
\expandafter\ifx\csname url\endcsname\relax
  \def\url#1{\texttt{#1}}\fi
\expandafter\ifx\csname urlprefix\endcsname\relax\def\urlprefix{URL }\fi
\expandafter\ifx\csname href\endcsname\relax
  \def\href#1#2{#2} \def\path#1{#1}\fi

\bibitem{Bishop2006}
C.~M. Bishop, \href{http://www.springer.com/us/book/9780387310732}{{Pattern
  Recognition and Machine Learning}}, 2006.
\newline\urlprefix\url{http://www.springer.com/us/book/9780387310732}

\bibitem{NIPS2012_4824}
A.~Krizhevsky, I.~Sutskever, G.~E. Hinton,
  \href{http://papers.nips.cc/paper/4824-imagenet-classification-with-deep-convolutional-neural-networks.pdf}{{ImageNet
  Classification with Deep Convolutional Neural Networks}}, in: F.~Pereira,
  C.~J.~C. Burges, L.~Bottou, K.~Q. Weinberger (Eds.), Advances in Neural
  Information Processing Systems 25, MIT Press, 2012, pp. 1097--1105.
\newline\urlprefix\url{http://papers.nips.cc/paper/4824-imagenet-classification-with-deep-convolutional-neural-networks.pdf}

\bibitem{Lecun2015}
Y.~LeCun, Y.~Bengio, G.~Hinton,
  \href{http://www.nature.com/articles/nature14539}{{Deep learning}}, Nature
  521~(7553) (2015) 436--444.
\newblock \href {https://doi.org/10.1038/nature14539}
  {\path{doi:10.1038/nature14539}}.
\newline\urlprefix\url{http://www.nature.com/articles/nature14539}

\bibitem{jannach2010recommender}
D.~Jannach, M.~Zanker, A.~Felfernig, G.~Friedrich, {Recommender Systems: An
  Introduction}, Cambridge University Press, 2010.

\bibitem{zhang2019deep}
S.~Zhang, L.~Yao, A.~Sun, Y.~Tay, {Deep learning based recommender system: A
  survey and new perspectives}, ACM Computing Surveys (CSUR) 52~(1) (2019)
  1--38.

\bibitem{graves2013speech}
A.~Graves, A.-r. Mohamed, G.~Hinton, {Speech recognition with deep recurrent
  neural networks}, in: 2013 IEEE International Conference on Acoustics, Speech
  and Signal Processing, IEEE, 2013, pp. 6645--6649.

\bibitem{bojarski2016end}
M.~Bojarski, D.~{Del Testa}, D.~Dworakowski, B.~Firner, B.~Flepp, P.~Goyal,
  L.~D. Jackel, M.~Monfort, U.~Muller, J.~Zhang, Others, {End to end learning
  for self-driving cars}, arXiv preprint arXiv:1604.07316 (2016).

\bibitem{miotto2018deep}
R.~Miotto, F.~Wang, S.~Wang, X.~Jiang, J.~T. Dudley, {Deep learning for
  healthcare: review, opportunities and challenges}, Briefings in
  Bioinformatics 19~(6) (2018) 1236--1246.

\bibitem{Goodfellow2016}
I.~Goodfellow, Y.~Bengio, A.~Courville,
  \href{https://www.deeplearningbook.org}{{Deep Learning}}, MIT Press, 2016.
\newline\urlprefix\url{https://www.deeplearningbook.org}

\bibitem{Kong2018}
Q.~Kong, D.~T. Trugman, Z.~E. Ross, M.~J. Bianco, B.~J. Meade, P.~Gerstoft,
  {Machine learning in seismology: turning data into insights}, Seismological
  Research Letters 90~(1) (2018) 3--14.
\newblock \href {https://doi.org/10.1785/0220180259}
  {\path{doi:10.1785/0220180259}}.

\bibitem{Ross2019}
Z.~E. Ross, D.~T. Trugman, E.~Hauksson, P.~M. Shearer, {Searching for hidden
  earthquakes in Southern California}, Science 771~(May) (2019) 767--771.
\newblock \href {https://doi.org/10.1126/science.aaw6888}
  {\path{doi:10.1126/science.aaw6888}}.

\bibitem{Bergen2019}
K.~J. Bergen, P.~A. Johnson, M.~V. de~Hoop, G.~C. Beroza,
  \href{http://www.sciencemag.org/lookup/doi/10.1126/science.aau0323}{{Machine
  learning for data-driven discovery in solid Earth geoscience}}, Science
  363~(6433) (2019) eaau0323.
\newblock \href {https://doi.org/10.1126/science.aau0323}
  {\path{doi:10.1126/science.aau0323}}.
\newline\urlprefix\url{http://www.sciencemag.org/lookup/doi/10.1126/science.aau0323}

\bibitem{Brenner2019}
M.~P. Brenner, J.~D. Eldredge, J.~B. Freund,
  \href{https://doi.org/10.1103/PhysRevFluids.4.100501}{{Perspective on machine
  learning for advancing fluid mechanics}}, Physical Review Fluids 4~(10)
  (2019) 100501.
\newblock \href {https://doi.org/10.1103/PhysRevFluids.4.100501}
  {\path{doi:10.1103/PhysRevFluids.4.100501}}.
\newline\urlprefix\url{https://doi.org/10.1103/PhysRevFluids.4.100501}

\bibitem{Brunton2020}
S.~L. Brunton, B.~R. Noack, P.~Koumoutsakos, {Machine Learning for Fluid
  Mechanics}, Annual Review of Fluid Mechanics 52~(1) (2020) 477--508.
\newblock \href {http://arxiv.org/abs/1905.11075} {\path{arXiv:1905.11075}},
  \href {https://doi.org/10.1146/annurev-fluid-010719-060214}
  {\path{doi:10.1146/annurev-fluid-010719-060214}}.

\bibitem{Dana2020}
S.~Dana, M.~F. Wheeler, \href{http://arxiv.org/abs/2003.11372}{{A machine
  learning accelerated $\textrm{FE}^2$ homogenization algorithm for elastic
  solids}} (2020).
\newblock \href {http://arxiv.org/abs/2003.11372} {\path{arXiv:2003.11372}}.
\newline\urlprefix\url{http://arxiv.org/abs/2003.11372}

\bibitem{Tartakovsky2018}
A.~M. Tartakovsky, C.~O. Marrero, P.~Perdikaris, G.~D. Tartakovsky,
  D.~Barajas-Solano, \href{http://arxiv.org/abs/1808.03398}{{Learning
  Parameters and Constitutive Relationships with Physics Informed Deep Neural
  Networks}} (2018).
\newblock \href {http://arxiv.org/abs/1808.03398} {\path{arXiv:1808.03398}}.
\newline\urlprefix\url{http://arxiv.org/abs/1808.03398}

\bibitem{Xu2020}
K.~Xu, D.~Z. Huang, E.~Darve, \href{http://arxiv.org/abs/2004.00265}{{Learning
  Constitutive Relations using Symmetric Positive Definite Neural Networks}}
  (2020) 1--31\href {http://arxiv.org/abs/2004.00265}
  {\path{arXiv:2004.00265}}.
\newline\urlprefix\url{http://arxiv.org/abs/2004.00265}

\bibitem{Raissi2019}
M.~Raissi, P.~Perdikaris, G.~E. Karniadakis,
  \href{https://doi.org/10.1016/j.jcp.2018.10.045}{{Physics-informed neural
  networks: A deep learning framework for solving forward and inverse problems
  involving nonlinear partial differential equations}}, Journal of
  Computational Physics 378 (2019) 686--707.
\newblock \href {https://doi.org/10.1016/j.jcp.2018.10.045}
  {\path{doi:10.1016/j.jcp.2018.10.045}}.
\newline\urlprefix\url{https://doi.org/10.1016/j.jcp.2018.10.045}

\bibitem{Raissi2018d}
M.~Raissi, A.~Yazdani, G.~E. Karniadakis,
  \href{http://arxiv.org/abs/1808.04327}{{Hidden Fluid Mechanics: A
  Navier-Stokes Informed Deep Learning Framework for Assimilating Flow
  Visualization Data}} (2018).
\newblock \href {http://arxiv.org/abs/1808.04327} {\path{arXiv:1808.04327}}.
\newline\urlprefix\url{http://arxiv.org/abs/1808.04327}

\bibitem{Haghighat2020}
E.~Haghighat, M.~Raissi, A.~Moure, H.~Gomez, R.~Juanes,
  \href{http://arxiv.org/abs/2003.02751}{{A deep learning framework for
  solution and discovery in solid mechanics}} (2020).
\newblock \href {http://arxiv.org/abs/2003.02751} {\path{arXiv:2003.02751}}.
\newline\urlprefix\url{http://arxiv.org/abs/2003.02751}

\bibitem{Rudy2019}
S.~Rudy, A.~Alla, S.~L. Brunton, J.~N. Kutz,
  \href{https://epubs.siam.org/doi/10.1137/18M1191944}{{Data-Driven
  Identification of Parametric Partial Differential Equations}}, SIAM Journal
  on Applied Dynamical Systems 18~(2) (2019) 643--660.
\newblock \href {https://doi.org/10.1137/18M1191944}
  {\path{doi:10.1137/18M1191944}}.
\newline\urlprefix\url{https://epubs.siam.org/doi/10.1137/18M1191944}

\bibitem{bergstra2010theano}
J.~Bergstra, O.~Breuleux, F.~Bastien, P.~Lamblin, R.~Pascanu, G.~Desjardins,
  J.~Turian, D.~Warde-Farley, Y.~Bengio, {Theano: a CPU and GPU math expression
  compiler}, in: Proceedings of the Python for Scientific Computing Conference
  (SciPy), Vol.~4, Austin, TX, 2010.

\bibitem{abadi2016tensorflow}
M.~Abadi, P.~Barham, J.~Chen, Z.~Chen, A.~Davis, J.~Dean, M.~Devin,
  S.~Ghemawat, G.~Irving, M.~Isard, M.~Kudlur, J.~Levenberg, R.~Monga,
  S.~Moore, D.~G. Murray, B.~Steiner, P.~Tucker, V.~Vasudevan, P.~Warden,
  M.~Wicke, Y.~Yu, X.~Zheng,
  \href{https://www.usenix.org/conference/osdi16/technical-sessions/presentation/abadi}{{TensorFlow:
  A system for large-scale machine learning}}, in: 12th USENIX Symposium on
  Operating Systems Design and Implementation (OSDI 16), USENIX Association,
  Savannah, GA, 2016, pp. 265--283.
\newline\urlprefix\url{https://www.usenix.org/conference/osdi16/technical-sessions/presentation/abadi}

\bibitem{chen2015mxnet}
T.~Chen, M.~Li, Y.~Li, M.~Lin, N.~Wang, M.~Wang, T.~Xiao, B.~Xu, C.~Zhang,
  Z.~Zhang, {MXNet: A flexible and efficient machine learning library for
  heterogeneous distributed systems} (2015).
\newblock \href {http://arxiv.org/abs/1512.01274} {\path{arXiv:1512.01274}}.

\bibitem{chollet2015keras}
F.~Chollet, \href{https://books.google.ca/books?id=Yo3CAQAACAAJ}{Deep Learning
  with Python}, Manning Publications Company, 2017.
\newline\urlprefix\url{https://books.google.ca/books?id=Yo3CAQAACAAJ}

\bibitem{Gune2018}
A.~G{\"{u}}ne¸, G.~Baydin, B.~A. Pearlmutter, J.~M. Siskind,
  \href{http://www.jmlr.org/papers/volume18/17-468/17-468.pdf}{{Automatic
  Differentiation in Machine Learning: a Survey}}, Journal of Machine Learning
  Research 18 (2018) 1--43.
\newline\urlprefix\url{http://www.jmlr.org/papers/volume18/17-468/17-468.pdf}

\bibitem{Kharazmi2020}
E.~Kharazmi, Z.~Zhang, G.~E. Karniadakis,
  \href{http://arxiv.org/abs/2003.05385}{{hp-VPINNs: Variational
  Physics-Informed Neural Networks With Domain Decomposition}} (2020)
  1--21\href {http://arxiv.org/abs/2003.05385} {\path{arXiv:2003.05385}}.
\newline\urlprefix\url{http://arxiv.org/abs/2003.05385}

\bibitem{Hornik1989}
K.~Hornik, M.~Stinchcombe, H.~White, {Multilayer feed-forward networks are
  universal approximators}, Neural Networks 2~(5) (1989) 359--366.
\newblock \href {https://doi.org/10.1016/0893-6080(89)90020-8}
  {\path{doi:10.1016/0893-6080(89)90020-8}}.

\bibitem{Cybenko1989}
G.~Cybenko, \href{https://doi.org/10.1007/BF02551274}{{Approximation by
  superpositions of a sigmoidal function}}, Mathematics of Control, Signals and
  Systems 2~(4) (1989) 303--314.
\newblock \href {https://doi.org/10.1007/BF02551274}
  {\path{doi:10.1007/BF02551274}}.
\newline\urlprefix\url{https://doi.org/10.1007/BF02551274}

\bibitem{Hornik1991}
K.~Hornik,
  \href{http://www.sciencedirect.com/science/article/pii/089360809190009T}{Approximation
  capabilities of multilayer feed-forward networks}, Neural Networks 4~(2)
  (1991) 251 -- 257.
\newblock \href {https://doi.org/https://doi.org/10.1016/0893-6080(91)90009-T}
  {\path{doi:https://doi.org/10.1016/0893-6080(91)90009-T}}.
\newline\urlprefix\url{http://www.sciencedirect.com/science/article/pii/089360809190009T}

\bibitem{Rumelhart1986}
D.~E. Rumelhart, G.~E. Hinton, R.~J. Williams, {Learning representations by
  back-propagating errors}, Nature 323~(6088) (1986) 533--536.
\newblock \href {https://doi.org/10.1038/323533a0}
  {\path{doi:10.1038/323533a0}}.

\bibitem{dafermos-claws}
C.~M. Dafermos, Hyperbolic Conservation Laws in Continuum Physics,
  Springer-Verlag, Berlin, 2000.

\bibitem{Raissi2018}
M.~Raissi, {Deep hidden physics models: Deep learning of nonlinear partial
  differential equations}, Journal of Machine Learning Research 19 (2018)
  1--24.
\newblock \href {http://arxiv.org/abs/1801.06637} {\path{arXiv:1801.06637}}.

\bibitem{simohughes-ci}
J.~C. Simo, T.~J.~R. Hughes, Computational Inelasticity, Vol.~7 of
  Interdisciplinary Applied Mathematics, Springer, New York, 1998.

\bibitem{zienkiewicz1969elasto}
O.~Zienkiewicz, S.~Valliappan, I.~King, Elasto-plastic solutions of engineering
  problems ‘initial stress’, finite element approach, International Journal
  for Numerical Methods in Engineering 1~(1) (1969) 75--100.

\bibitem{COMSOL}
{COMSOL}, {COMSOL} {M}ultiphysics User's Guide, {COMSOL}, Stockholm, Sweden,
  2020.

\bibitem{Weinan2018}
E.~Weinan, B.~Yu, {The Deep Ritz Method: A Deep Learning-Based Numerical
  Algorithm for Solving Variational Problems}, Communications in Mathematics
  and Statistics 6~(1) (2018) 1--14.
\newblock \href {http://arxiv.org/abs/1710.00211} {\path{arXiv:1710.00211}},
  \href {https://doi.org/10.1007/s40304-018-0127-z}
  {\path{doi:10.1007/s40304-018-0127-z}}.

\bibitem{Berg2018}
J.~Berg, K.~Nystr{\"{o}}m, \href{https://doi.org/10.1016/j.neucom.2018.06.056
  https://linkinghub.elsevier.com/retrieve/pii/S092523121830794X}{{A unified
  deep artificial neural network approach to partial differential equations in
  complex geometries}}, Neurocomputing 317 (2018) 28--41.
\newblock \href {http://arxiv.org/abs/1711.06464} {\path{arXiv:1711.06464}},
  \href {https://doi.org/10.1016/j.neucom.2018.06.056}
  {\path{doi:10.1016/j.neucom.2018.06.056}}.
\newline\urlprefix\url{https://doi.org/10.1016/j.neucom.2018.06.056
  https://linkinghub.elsevier.com/retrieve/pii/S092523121830794X}

\end{thebibliography}

\end{document}